\documentstyle[prl,aps,multicol]{revtex}

\begin{document}
\draft
\preprint{}
\title{Teleportation as a Depolarizing Quantum Channel, Relative Entropy and Classical Capacity}
\author{Garry Bowen \cite{email} and Sougato Bose}
\address{Centre for Quantum Computation, Clarendon Laboratory, University of Oxford, Oxford OX1 3PU, United Kingdom.}
\date{\today}
\maketitle

\begin{abstract}
We show that standard teleportation with an arbitrary mixed state
resource is equivalent to a generalized depolarizing channel with
probabilities given by the maximally entangled components of the
resource. This enables the usage of any quantum channel as a
generalized depolarizing channel without additional twirling
operations. It also provides a nontrivial upper bound on the
entanglement of a class of mixed states. Our result allows a
consistent and statistically motivated quantification of
teleportation success in terms of the relative entropy and this
quantification can be related to a classical capacity.
\end{abstract}

\begin{multicols}{2}


The possibility of transferring an unknown quantum state using
pre-existing entanglement and a classical information channel was
labeled {\it teleportation} by its authors \cite{bennett93}. The
teleportation process can be viewed as a quantum channel. The
nature of the channel is determined by both the state used as a
teleportation resource, and the particular protocol used with
this resource \cite{horodecki99,horodecki96a,dariano00}. The standard 
teleportation protocol $T_0$ using the Bell diagonal measurements
and Pauli rotations, when used in conjunction with a Bell state
resource, provides an example of a {\em noiseless} quantum channel
$\Lambda_{T_0}(|\Psi^+\rangle \langle \Psi^+ |)\varrho =
\varrho$. Teleportation using mixed states as an entanglement
resource is, in general, equivalent to a {\em
noisy} quantum channel. A general 
expression for the {\em output state} of a teleportation process
with an {\em arbitrary} mixed resource, in terms of some quantum          
channel, has been shown previously \cite{dariano00}. In this
letter, we derive an explicit expression for the quantum channel
associated with the standard
teleportation protocol on a mixed state resource. 
Our result establishes a many to one
correspondence between arbitrary bipartite quantum states and
generalized depolarizing channels. This is a complete
generalization of an earlier correspondence noted by the
Horodecki's \cite{horodecki99} between quantum channels $\Lambda$
and the {\em restricted class} of quantum states $\rho_{\Lambda}$
with one reduced density matrix equal to the maximally mixed
state. We then present both practical and theoretical
applications of our result. From a practical point of view, our
result allows an arbitrary quantum channel to be used as a
generalized depolarizing channel. It permits Bell diagonal states
to be shared between ends of an arbitrary channel without resort
to the time consuming twirling operations
\cite{horodecki99,bennett96b}. On the theoretical side, our
result can be used to obtain a {\em nontrivial} upper bound on
the entanglement of a certain class of mixed states. We then show
that our result allows the quantification of the success of
teleportation consistently (i.e. without any divergence) in terms
of the relative entropy. This quantification {\em unifies}
 the methodology of quantification of teleportation
success with that of entanglement \cite{vedral01,vedral98} and
classical capacity
\cite{holevo98,bose00,bowen01,hiroshima01,horodecki00}. More
importantly, it gives a statistical interpretation of the
teleportation success, with {\em collective measurements} being
allowed on an ensemble of $N$ teleported states after $N$ separate
teleportation processes \cite{vedral01}. The currently used {\em
fidelity} of teleportation \cite{horodecki99} fails to do this. In
the end we show that teleportation success, as quantified by the
relative entropy, is bounded above by a classical capacity. This
relation can be regarded as connecting a quantum and a classical
capacity.

 We start by stating the main result of the letter before going into its proof.
 It states that the standard teleportation
protocol $T_0$, when used with an arbitrary two qubit mixed state,
$\chi$, as a resource, acts as a generalized depolarizing channel,
\begin{equation}
\Lambda_{T_0}(\chi)\varrho = \sum_i {\mathrm Tr}[E^i \chi] \,
\sigma^i \varrho \sigma ^i , \label{eqn:channel}
\end{equation}
where the $E^i$'s are the Bell states associated with the Pauli
matrices $\sigma^i$, by $E^i = \sigma^i E^0 \sigma^i$, where $E^0
= |\Psi^+ \rangle \langle \Psi^+|$ and $\sigma^0 = I,\sigma^1 =
\sigma_x, \sigma^2 = \sigma_y$ and $\sigma^3 = \sigma_z$.

The result generalizes the relationship between particular
teleportation protocols and quantum channels to include {\em all}
$2 \times 2$ mixed states, and proves the conjecture (made in
Ref.\cite{horodecki99}) that the relationship between mixed
states used for teleportation and the resultant quantum channel
is not one to one.  The derivation, it may be noted, rests
critically on the linearity of the teleportation protocol
\cite{bennett93}.  We will also extend the result to
teleportation with  $d \times d$ state systems. We next proceed
to the derivation of our central result.

Suppose Alice wishes to teleport the unknown qubit $\varrho$,
then initially we can extend this state to a $2 \times 2$ pure
state $|\psi \rangle_{12}$, even if $\varrho$ is initially pure,
such that ${\mathrm Tr}_{2} [ |\psi \rangle \langle \psi |_{12} ]
= \varrho$.  We then teleport only the original state $\varrho$,
and examine the outcome by comparing the total state $|\psi
\rangle$ to the entanglement swapped state. Since an arbitrary
state in a $2 \times 2$ system may be written in terms of a
superposition of Bell basis states,
\begin{equation}
|\psi \rangle = c_0 |\Psi^+ \rangle + c_1 |\Psi^- \rangle + c_2 |\Phi^+ \rangle + c_3 |\Phi^- \rangle ,
\end{equation}
and because of the linearity of the teleportation protocol, we need only look at how the component Bell states of the density matrix $|\psi \rangle \langle \psi |$ are affected by teleportation using the $T_0$ protocol, using an arbitrary resource $\rho$.

We label the $2 \times 2$ state used in the teleportation by
$|\psi \rangle_{12}$ and the resource by $\chi_{34}$, where the
subscripts denote the particle number.  Alice has qubits in states
$\chi_{3}={\mathrm Tr}_{4}[\chi_{34}]$ and $\varrho_{1} = {\mathrm
Tr}_{2}[|\psi \rangle \langle \psi |_{12} ]$, and Bob has a qubit
in the state $\chi_{4}={\mathrm Tr}_{3}[\chi_{34}]$.  The outcome
of the teleportation is the state,
\begin{equation}
\Lambda_{T_0}(\chi_{34}) |\psi \rangle \langle \psi |_{12} =
\omega_{24}. \label{eqn:output}
\end{equation}
Choosing the basis state $|\psi \rangle_{12} = |\Psi^+ \rangle$,
in Eq.(\ref{eqn:output}), we note that the teleportation then
becomes a version of entanglement swapping
\cite{bennett93,entswp1,entswp2} with one perfect and one noisy
entangled state. Given a measurement outcome of the $i$-th state
upon measurement, we know that the final state, before the unitary
operation, is in the state, $\omega^{i}_{24} = \sigma_{2}^{i}
\chi_{24} \sigma_{2}^{i}$, because this is equivalent to
teleportation with the state $|\Psi^{+}\rangle_{12}$, without
applying the unitary transform $\sigma_{2}^{i}$ to the output
state, and $\sigma^i = (\sigma^i)^{\dag} = (\sigma^i)^{-1}$.

As the teleportation uses the channel $\chi_{34}$, the unitary
operation is applied to $\chi_{4}$, and the output state is then,
\begin{equation}
\omega^{i}_{24} = \sigma_{4}^{i} \sigma_{2}^{i} \chi_{24}
\sigma_{2}^{i} \sigma_{4}^{i} , \label{eqn:teleport-i}
\end{equation}
and therefore, over all outcomes $i$, the final total teleported state is,
\begin{equation}
\omega_{24} = \sum_i p_i \, \omega^{i}_{24} = \sum_i p_i \,
\sigma_{4}^{i} \sigma_{2}^{i} \chi_{24} \sigma_{2}^{i}
\sigma_{4}^{i} ,
\end{equation}
where $p_i$ is the chance of obtaining outcome $i$ upon measurement.

A tedious calculation will show that the probability of gaining outcome $i$, for the combined Bell state measurements on qubits 1 and 3, is simply $p_i = 1/4$.  Hence, we can move the summation to obtain,
\begin{eqnarray}
\omega_{24} &=& \frac{1}{4} \sum_i \sigma_{4}^{i} \sigma_{2}^{i} \chi_{24} \sigma_{2}^{i} \sigma_{4}^{i} \label{eqn:main1} \\
&=& \sum_i E_{24}^i \chi_{24} E_{24}^i \label{eqn:main2} \\
&=& \sum_i {\mathrm Tr}[ E_{34}^i \chi_{34}] \, E_{24}^i \\
&=& \sum_i {\mathrm Tr}[ E_{34}^i \chi_{34}] \, \sigma_{4}^i
|\Psi^{+} \rangle \langle \Psi^{+}|_{24} \sigma_{4}^{i} .
\label{eqn:main4}
\end{eqnarray}
The equality between Eq.(\ref{eqn:main1}) and Eq.(\ref{eqn:main2}) can be shown by decomposing the Pauli operators in Eq.(\ref{eqn:main1}) in terms of the Bell state projectors, for example, $\sigma^1_2 \sigma^1_4 = E^0_{24}+E^1_{24}-E^2_{24}-E^3_{24}$, and noting that all terms except those of the form given in Eq.(\ref{eqn:main2}) cancel.

Substituting another Bell state $E_{12}^{j}$ into Eq. (\ref{eqn:output}) simply rotates the output state by the corresponding Pauli operator $\sigma^{j}_{2} \omega^{i}_{24} \sigma_{2}^{j}$, and so,
\begin{eqnarray}
\omega^{(j)}_{24} &=& \sigma_{2}^{j} \big( \sum_i {\mathrm Tr}[ E_{34}^i \chi_{34}] \, E_{24}^i \big) \sigma_{2}^{j} \\
&=& \sum_i {\mathrm Tr}[ E_{34}^i \chi_{34}] \, \sigma_{4}^i
E^{j}_{24} \sigma_{4}^{i} .
\end{eqnarray}
Additionally, the off diagonal Bell terms, $F^{mn} = \sigma^{m}
|\Psi^+ \rangle \langle \Psi^+ | \sigma^{n}$, for $m\neq n$,
follow by the linearity of the standard teleportation protocol,
\begin{eqnarray}
\omega^{(mn)}_{24} &=& \sigma^{m}_{2} \big( \sum_i {\mathrm Tr}[ E_{34}^i \chi_{34}] \, E_{24}^i \big) \sigma_{2}^{n} \\
&=& \sum_i {\mathrm Tr}[ E_{34}^i \chi_{34}] \, \sigma_{4}^i
F^{mn}_{24} \sigma_{4}^{i} .
\end{eqnarray}
The total final teleported state, given an arbitrary state $|\psi
\rangle \langle \psi |_{12}$ as input, is then,
\begin{eqnarray}
\omega_{24} &=&  |c_0|^2 \sum_i {\mathrm Tr}[ E_{34}^i \chi_{34}] \, \sigma_{4}^i |\Psi^{+} \rangle \langle \Psi^{+}|_{24} \sigma_{4}^{i} \nonumber \\
&& \quad + \sum_{j \neq 0} |c_j|^2 \sum_i {\mathrm Tr}[ E_{34}^i \chi_{34}] \, \sigma_{4}^i E^{j}_{24} \sigma_{4}^{i} \nonumber \\
&& \quad \quad + \sum_{m\neq n} c_m c_n^* \sum_i {\mathrm Tr}[ E_{34}^i \chi_{34}] \, \sigma_{4}^i F^{mn}_{24} \sigma_{4}^{i} \\
&=& \sum_i {\mathrm Tr}[ E_{34}^i \chi_{34}] \, \sigma_{4}^i |\psi
\rangle \langle \psi |_{24} \sigma_{4}^{i} , \label{eqn:result}
\end{eqnarray}
and by tracing over qubit 2 in Eq.(\ref{eqn:result}) and comparing with Eq.(\ref{eqn:channel}) we can see that the channel acts as a generalized depolarization channel,
\begin{equation}
\Lambda_{T_0}(\chi)\varrho = \sum_i p_i \, \sigma^i \varrho
\sigma^i ,
\label{telpout}
\end{equation}
with the probabilities given by the projections of the Bell
states on the teleportation resource $p_i = \mathrm{Tr}[ E^i
\chi]$.
The above result has been proved so far only for the
teleportation of state $\rho$ of a single qubit.  From
Eq.(\ref{eqn:result}) and {\em linearity}, it can easily be
extended to the case of teleportation of one half of a $2\times2$
mixed state $\gamma_{12}$ (i.e. for entanglement swapping)
through a bipartite resource $\chi_{34}$. We will simply have to
replace $|\psi \rangle \langle \psi |_{24}$ in
Eq.(\ref{eqn:result}) by $\gamma_{24}$ in order to obtain the
output state of the teleportation process. Teleportation of an
entangled mixed state $\gamma_{12}$ is thus given by

\begin{equation}
\Lambda_{T_0}(\chi)\gamma = \sum_i {\mathrm Tr}[E^i \chi] \,
\sigma_{4}^i \gamma_{24}\sigma_{4}^i. \label{entswpout}
\end{equation}
Eqs.(\ref{telpout}) and (\ref{entswpout}) are the first ever
general expressions for teleportation and entanglement swapping
with {\em arbitrary} mixed states, as long as the teleportation
protocol is kept standard. One must remember that for optimal
utilization of a given entangled resource, one must choose local
basis states such that $p_0$ is maximum. One can regard this
particular state as the principal state ($|\Psi^{+}\rangle$) of
the teleportation protocol. In principle, it should allow one to
rederive all known results about the standard protocol (for
example, the dependence of teleportation fidelity on the
maximally entangled fraction \cite{horodecki99} only). However,
in the rest of this letter, we will explore those consequences of
our result which are unknown to date.

  Eq.(\ref{entswpout}) immediately provides an upper bound to the
entanglement of a class of mixed states. From the fact that
entanglement cannot be increased under local actions and
classical communications, it follows that the output entangled
state $\lambda=\Lambda_{T_0}(\chi)\gamma$ must have an
entanglement lower than that of the less entangled of the states
$\gamma_{12}$ and $\chi_{34}$. Therefore, for any state
$\lambda_{12}$ expressible in terms of another state
$\gamma_{12}$ as $\sum_i p_i \sigma_{2}^i \gamma_{12}
\sigma_{2}^i$, the entanglement
\begin{equation}
{\mathcal E}(\lambda) \leq  {\mathcal E}(\beta\{p_i\})\; ,
\label{eqn:ent_bound}
\end{equation}
where, $\beta\{p_i\}$ denotes the class of states with
Bell-diagonal projections $p_i$. This bound implies that for
generlized qubit depolarizing channels, $\Lambda$, with a
spectrum $p_i \in [0,1/2]$, we have $\Lambda\rho$ to be separable
for all $\rho$. In other words, no matter what initial state you
use, you can {\em never} establish entanglement between the ends
of such a channel. When $\beta\{p_i\}$ are taken to be Bell
diagonal states, the {\em upper} bound of
Eq.(\ref{eqn:ent_bound}) will {\em complement} the usual {\em
lower} bounds on entanglement of states obtained by wernerization
\cite{bennett96b}. The above bound implies that the entanglement
left after passing one half of an arbitrary mixed entangled state
$\gamma$ through a generalized depolarizing channel is less than
or equal to that left when one half of a Bell state is passed
through the channel. The {\em non-triviality} of the result stems
from the fact that even if $\gamma$ is obtainable from a Bell
state by action of local operators, these operators do not
necessarily commute with those of the depolarizing channel.

   The next noteworthy consequence of our result is that it provides an alternative to the use of
time consuming twirling operations \cite{horodecki99,bennett96b}
in quantum communication protocols. Such operations involve
applying random local unitary operations to an entangled pair of
particles to bring them to a Bell diagonal state. Here, firstly,
there is the problem of the choice of local operations (being
decided clasically) being pseudo-random. Secondly, it has to be
done to a large enough ensemble, and later on, the memory of which
random rotation was applied to which pair has to be forgotten.
Obtaining Bell-diagonal states via twirling could thus
potentially be a very time consuming process. Our result,
Eq.(\ref{entswpout}), clearly illustrates that one can produce a
Bell-diagonal state from any mixed state by local actions {\em
without twirling}. One simply has to teleport the state
$|\psi^{+}\rangle$ through the given mixed state using the
standard teleportation protocol.  Each member of the resultant
ensemble is already in a Bell diagonal state without the
necessity of forgetting any local actions. Moreover, the
randomness is intrinsic ``quantum'' randomness, stemming from the
teleportation protocol.

  We pause here briefly to provide the generalization of our derivation to
higher dimensional analogues of the standard teleportation
scheme, with mixed resource states. For a $d \times d$ state
system, the standard teleportation scheme is constructed using
the maximally entangled state, $|\Psi^+ \rangle =
\frac{1}{\sqrt{d}} \sum_j |j \rangle |j \rangle$, and the set of
unitary generators $U_{(1)}^{nm} = \sum_k e^{2\pi i k n/d} |k
\rangle \langle k \oplus m |$, acting on the first part of the
system, where $\oplus$ denotes addition modulo-$d$.  The set of
maximally entangled states are then denoted by $E^{nm} = U^{nm}
|\Psi^+ \rangle \langle \Psi^+ | (U^{nm})^{\dag}$, respectively,
for $n,m = 0,1,..., d-1$. If steps corresponding to those of
Eq.(\ref{eqn:teleport-i}) to Eq.(\ref{eqn:main4}) is carefully
carried out in this case, the higher dimensional teleportation
channel remains a depolarizing channel of the form,
\begin{equation}
\Lambda \varrho = \sum_{nm} {\mathrm Tr}[E^{nm} \rho] \,
U^{n(-m)} \varrho (U^{n(-m)})^{\dag} . \label{eqn:extension}
\end{equation}

    Now we proceed to one of the most important consequences of our result, namely, the fact that the teleported state
(Eqs.(\ref{telpout}),(\ref{entswpout}) and (\ref{eqn:extension}))
     is {\em always mixed}, apart
  from the isolated case of maximally entangled channel. This implies that
  the relative entropy between the input state and the output state  will always be
  {\em finite}. This allows us to quantify the success of teleportation
  using the relative entropy. Without our result
  (Eqs.(\ref{telpout}) and (\ref{eqn:extension})), there is no way
  to be sure that relative entropy between the input and the
  output state of the standard teleportation protocol would not
  blow up.
The quantum relative entropy
\cite{vedral01,vedral98,schumacher00} is defined as $S( \rho ||
\omega ) = {\mathrm Tr}\, [ \rho \log \rho - \rho \log \omega]$
and has a statistical interpretation \cite{vedral97}, where the
probability of mistaking the state $\omega$ for the state $\rho$
after $N$ measurements is given by $P(\omega \rightarrow \rho)
\simeq e^{-N S(\rho || \omega)}$ as $N \rightarrow \infty$. The
success of teleportation may then be given by,
\begin{equation}
{\mathcal F} = \overline{S(\psi_{in} || \omega_{out} )} ,
\label{eqn:relinout}
\end{equation}
averaged over all pure input states, $\psi_{in}$, in a similar
way to fidelity, and $\omega_{out}$ is the output state.
Physically, this has significance when a third party wishes to
verify a, possibly imperfect, teleportation between two untrusted
parties. We define imperfect as meaning the teleporting parties
share no entanglement.  The probability of the third party being
fooled by the imperfect teleportation scheme, for a large number
of states $N$, is given by $e^{-N S( \psi_{in}|| \omega_{out})}$,
even assuming the third party is making optimal generalized collective 
measurements over the $N$ teleportations. The relative entropy
thus provides an asymptotic (collective) measure of teleportation
success compared to the ``single shot'' nature of the fidelity
measure.

 The above measure can be readily applied to demonstrate that teleporting one half of
a maximally entangled state is a better way to detect the
presence of entanglement than teleporting a single $d$ state
system. Using the quantum relative entropy to examine the fidelity
of entanglement swapping, we can choose the state $|\Psi^+
\rangle$ as the input state, and the relative entropy, ${\mathcal
F}^+$, is given by,
\begin{eqnarray}
S\big(|\Psi^+ \rangle \langle \Psi^+ |  \big|\big| \, \omega_{out} \big) &=& - {\mathrm Tr} \big[ |\Psi^+ \rangle \langle \Psi^+ | \log \omega_{out} \big] \\
&=& -\log {\mathrm Tr} \big[|\Psi^+ \rangle \langle \Psi^+ | \, \rho \big] \\
&=& -\log F ,
\end{eqnarray}
the negative log of the singlet fraction of the resource $\rho$.
The maximally entangled fraction for separable states is bound by,
$1/d^2 \leq F \leq 1/d$, which gives bounds on the relative
entropy,
\begin{equation}
\log d^2 \geq {\mathcal F}^+ \geq \log d .
\end{equation}
Since Eq.(\ref{eqn:relinout}) is bounded above by, ${\mathcal F}
\leq \log d$ (for teleportation of a single $d$ state system), the
optimal method for verification of the {\em presence of
entanglement} through teleportation is by sending half of a
maximally entangled $d\times d$ pair through the teleportation
channel.

   We now proceed to show how the success of teleportation, when
quantified by the relative entropy, can be related to a classical
capacity. The classical capacity of communication using the
quantum states $\rho_i=\sigma_i\rho\sigma_i$ as letters (For
qubits $\sigma_i$ are the Pauli matrices and identity, while for
higher dimensions, they are corresponding generalizations) , with
{\em apriori} probabilities $p_i$ is given \cite{holevo98} by
$C=\sum_i p_i S(\rho_i||\sum_j p_j\rho_j)$. From
Eqs.(\ref{telpout}) and (\ref{eqn:relinout}), it is clear that
each term in the above summation can be interpreted as an {\em
unaveraged} relative entropy measure of success of a standard
teleportation protocol with a different utilization of the same
resource. The particular state to be teleported is $\rho$ and the
resource $\chi$ has maximally entangled components with weights
$p_i$. While the first term corresponds to optimal utilization of
the resource for teleportation, the other three terms correspond
to a less efficient teleportation using the maximally entangled
components of lower weight as the principal state
($|\Psi^{+}\rangle$) for teleportation. Worse teleportation
implies greater value of the relative entropy between the input
and the output state, by virtue of which we have,
\begin{equation}
{\mathcal F} \leq \overline{C},
\end{equation}
where $\overline{C}$ is the average of $C$ taken over all possible
pure input $\rho=\psi_{in}$. Physically, this means that the
relative entropy measure of teleportation success will be bounded
above by the average classical communication capacity using pure
letter states related by Pauli rotations with {\em apriori
probabilities being given by the weights  of the maximally
entangled components of the resource}. This result can be regarded
as connecting a quantum and a classical capacity.

 In this letter we have presented an explicit expression for the output
 of a standard teleportation protocol using an arbitrary mixed
 resource. Most known results about the standard teleportation
 process \cite{horodecki99,bennett96b} follow quite
 straightforwardly from our expression. It also has the potential
 for generating a host of other results (of which, we have given three distinct examples) relating to the standard
 teleportation process with an arbitrary mixed state.
 Most importantly, our result {\em allows} us to define a statistical measure
 of teleportation success in terms of relative entropy. It will be straightforward
 to generalize our result to multi-party scenarios of entanglement swapping \cite{entswp2} with
 Greenberger-Horne-Zeilinger state measurements and arbitrary mixed states.
 The use of ``twisted'' entangled states
\cite{braunstein00} may also lead to the generalization of this
result to arbitrary teleportation schemes.

The authors would like to thank Konrad Banaszek for very helpful discussions.  GB is supported by the Harmsworth Trust, the Oxford-Australia Fund, and the CVCP.

\end{multicols}

\end{document}